\documentstyle[11pt]{article}
\begin{document}
\title{Chkareuli-Froggatt-Nielsen Theorem and Photon Mass}
\author{Haryanto M. Siahaan\footnote{E-mail : hm\_siahaan@gmail.com}\\}
\date{}
\maketitle

\begin{abstract}
We analyze there is a relation between fossil charge and the mass of photon based on 
Chkareuli-Froggatt-Nielsen Theorem and Proca Lagrangian. 
As generally known, massive photon will lead to Lorentz non-invariance field theory. 
\end{abstract}

\section{Introduction}
In this paper we review the theorem of Chkareuli-Froggatt-Nielsen (CFN) based on their paper in [1] first. Secondly, we will review on Proca Lagrangian which widely available on every QFT textbooks, for example [4]. Thirdly, we will make a relation between fossil charge density which appear in [1] and mass term in Proca Lagrangian. Finally, the discussion aout the result.

\section{CFN Theorem and Fossil Charge}
In their paper, [1], Chkareuli, Froggatt, and Nielsen start from a Lorentz invariant Lagrangian density plus a breaking term which only on $A_{\mu}$ via $A^{2}$. Then they assume that $A_{\mu}\left(x\right)$ takes on a non-zero constant value, providing a Lorentz symmetry breaking in vacuum. Universe is started with initial condition ensure that the partial derivative of the non-invariant part of Lagrangian to be vanish at early time, then the theory is Lorentz and gauge invariant with a non-covariant gauge choice properly fixed in the theory.
\begin{equation}
	\frac{\partial \tilde{L}_{br}}{\partial A^{2}}=0
\end{equation}
From the total Lagrangian density, $\tilde{L}=\tilde{L}_{kin}+\tilde{L}_{br}$, we can get an equation of motion which read as 
\begin{equation}
	\frac{\partial F^{\mu\nu}}{\partial x^{\mu}}=2A^{\nu}\frac{\partial \tilde{L}_{br}}{\partial\ A^{2}}+j^{\nu}.
\end{equation}
\par
They proved that the condition (1) will not change or always be fulfilled. But from condition (1) wo wont observe the Lorentz violation. So they made a small deviation from this condition by introducing approximate vacuum situation with a background vector field $A_{\mu}\approx V_{\mu}=n_{\mu}V$. So in general we can write the vector field current as
\begin{equation}
	j^{\mu}_{A}=2Vn^{\mu}\frac{\partial \tilde{L}_{br}}{\partial\ A^{2}},
\end{equation}
where $n_{\mu}$ is constant vector with $n^{2}=1$.

\section{Review on Proca Lagrangian Density}
Transition from Maxwell field to masive vector boson (Proca field) is easily to be performed by adding a quadratic mass term in Lagrangian density.
\begin{equation}
	\tilde{L}=-\frac{1}{4}F_{\mu\nu}F^{\mu\nu}+\frac{1}{2}m^{2}A_{\mu}A^{\mu}-j_{\mu}A^{\mu}
\end{equation}
Euler-Lagrange equation for (4)can be read as 
\begin{equation}
	\frac{\partial F^{\mu\nu}}{\partial x^{\mu}}+m^{2}A^{\nu}=j^{\nu}.
\end{equation}
Applying the Lorentz gauge\footnote{This means the theory is still Lorentz invariant} to equation (4), we arrive at Klein Gordon equation 
\begin{equation}
	\left(\partial_{\nu}\partial^{\nu}+m^{2}\right)A_{\mu}=j_{\mu}.
\end{equation}
We also can get the expression for energy and momentum density as
\begin{eqnarray}
	\tilde{E}=\frac{1}{2}\left(\vec{E}^{2}+\vec{B}^{2}\right)+\frac{1}{2}m^{2}\left(A_{0}+\vec{A}^{2}\right)-\vec{j}\cdot\vec{A},\\
	\tilde{\vec{p}}=\vec{E}\times\vec{B}-m^{2}A_{0}\vec{A}-j_{0}\vec{A}.
\end{eqnarray}

\section{Photon Mass and Fossil Charge}
From Proca Lagrangian density, we regard that the term which break Lorentz invariant is $\frac{1}{2}m^{2}A_{\mu}A^{\mu}$, so related to our notation before, we can write that
\begin{equation}
	\tilde{L}_{br}=\frac{1}{2}m^{2}A_{\mu}A^{\mu}.
\end{equation}
Based on equation (3), we also can make relation between photon mass, $m_{\gamma}$ (mass of Proca field), and "\textit{fossil resting charge density}" $\rho_{fossil}$ in time like case from equation (3) by writing
\begin{equation}
	\rho_{fossil}=Vm_{\gamma}^{2}.
\end{equation}
V can be regarded as a scalar potential (in time like case) that can be measured experimentally. According to some recent paper, for example [2] and [3], photon is viewed as massive particle, which means $\rho_{fossil}\neq 0$, and can be detected experimentally. Then as stated at their paper, [1], the existence of $\rho_{fossil}\neq 0$ is such a way for Lorentz non-invariance can come into theories.

\section{Discussion}
Although most of physicist still accept that photon is massless particle, there are also some who try to find the mass of photon. For example, as given in [2], the mass of photon has approximate upper value (in length unit), $m_{\gamma}\approx5\times 10^{-11}m$. This value is very small and because V is also small (perturbative), so $\rho_{fossil}$ will be very small too. At this point, if mass of photon is really not zero, it means that Lorentz symmetry is not valid anymore, and Chkareuli-Froggatt-Nielsen theorem seems able to accomodate it. Further explanation that needed is how the deviation from the initial condition (our universe is started with Lorentz invariant symmetry) could appear. This deviation of course will gives a small impact for energy and momentum density of photon related to Proca rather than standard Maxwell formulation.

\section{References}
\begin{enumerate}
	\item J. L. Chkareuli, C. D. Froggatt, H. B. Nielsen, hep-th/061018.
	\item L. B. Okun, Acta Phys. Pol. B 37, 565 (2006).
	\item R. Lakes, Phys. Rev. Lett. 80, 1826 (1998).
	\item W. Greiner, J. Reinhardt, "\textit{Field Quantization}," pp. 152, Springer-Verlag, 1996.
\end{enumerate}

\end{document}